\documentclass{PoS}
\usepackage{slashed}
\usepackage{amssymb}
\usepackage{amsmath}
\usepackage{amsfonts}

\newcommand{\be}{\begin{equation}}
\newcommand{\ee}{\end{equation}}
\newcommand{\bea}{\begin{eqnarray}}
\newcommand{\eea}{\end{eqnarray}}

\newcommand{\nn}{\nonumber\\}

\newcommand{\ol}{\overline}

\title{Some Novel Ways for Neutrino Mass Generation}

\ShortTitle{Unconventional Neutrino Mass Generation}

\author{\speaker{Nick E. Mavromatos}\thanks{Currently also at: CERN, Physics Department, Theory Division, Geneva 23 CH-1211, Switzerland.
This work is supported in part by the London Centre for Terauniverse Studies (LCTS), using funding from the European Research Council via 
the Advanced Investigator Grant 267352 and by STFC (UK) under the research grant ST/L000326/1.}\\
        King's College London, Physics Department, Theoretical Particle Physics and Cosmology Group,   Strand, London WC2R 2LS, UK\\
        E-mail: \email{Nikolaos.Mavromatos@cern.ch}}

\abstract{I discuss dynamical generation of neutrino masses in unconventional scenarios where the background space-time
geometry plays a crucial role. I discuss two types of backgrounds: (i) Lorentz Violating and (ii) Geometries with
Torsion. In the former case, the violation of Lorentz symmetry, at a scale M, may be viewed as a catalyst for mass
generation and induced flavour oscillations among neutrino species, which survive the limit of M taken to infinity,
leading to a hierarchy among neutrino masses. In the latter case, the (totally antisymmetric components of the) torsion degrees of freedom correspond to a
pseudoscalar axion field in four space-time dimensions. This field is assumed to be mixed, through non-diagonal kinetic terms, with 
ordinary axion fields that may exist in the theory for other reasons and
couple to neutrinos with chirality changing Yukawa couplings. The torsion-ordinary-axion-field mixing 
is responsible, through higher-loop anomalous graphs,
for the dynamical generation of Majorana masses. The latter scenarios can also be realised in some (compactified)
string theory models, where the (totally antisymmetric) torsion is played by the field strength of the spin-one antisymmetric tensor  (Kalb-Ramond) field, which exists in the gravitational string multiplet. }

\FullConference{XVI International Workshop on Neutrino Telescopes,\\
		2-6 March 2015\\
		Palazzo Franchetti †Istituto Veneto, Venice, Italy }

\makeatletter
\def\@oddfoot{\ifnum\thepage=1%
  \PoScopyright@box\hfill%
  \if@PoSspecialurl\PoSspecial@url\else\unhbox\PoSpaper@url\fi\fi}
\makeatother

\begin{document}

\section{Introduction and Summary}

The  discovery~\cite{higgs} of  the Higgs  boson at  the CERN
Large Hadron Collider (LHC) in 2012 constitutes an important milestone for the
Ultra-Violet (UV) completion of the Standard Model (SM).  Although the
so-called Higgs mechanism  may well explain the generation  of most of
the particle masses in the SM, the origin of the small neutrino masses
still remains an open issue.  In particular, the observed smallness of
the  light neutrino  masses  may naturally  be  explained through  the
see-saw  mechanism~\cite{seesaw},   which  necessitates  the  Majorana
nature of the light (active)  neutrinos and postulates the presence of
heavy right-handed  Majorana partners of mass  $M_R$. The right-handed
Majorana mass $M_R$  is usually considered to be  much larger than the
lepton or  quark masses.  The  origin of $M_R$  has been the  topic of
several extensions of  the SM in the literature,  within the framework
of quantum field  theory~\cite{seesaw,Schechter:1980gr}    
and string theory~\cite{Blumenhagen:2006xt}. However, up to now, there is no experimental evidence for right-handed neutrinos or for any extension of SM, as a matter of fact, although some optimism of discovering supersymmetry in the next round of LHC (operating at 14 TeV energies)
exists among particle physicists. 

Until therefore such extensions of the SM are discovered, it is legitimate to search for alternative mechanisms for neutrino mass generation, 
that keep the spectrum of SM intact, except perhaps for the existence of right handed neutrinos that are allowed. 
Such minimal, non supersymmetric extensions of the Standard Model with three in fact right-handed Majorana neutrinos complementing the three active 
left-handed neutrinos  (termed $\nu$MSM), have been proposed~\cite{nuMSM}, in a way consistent with current cosmology. Such models are characterised by relatively light right-handed neutrinos, two of which are almost degenerate, with masses of order GeV, and a much lighter one, almost decoupled, with masses in the keV range, which may play the role of warm dark matter. The right-handed neutrinos serve the purpose of generating, , through seesaw type mechanisms, the active neutrino mass spectrum, consistent with observed flavour oscillations. However, there are no suggestions for microscopic mechanisms for the generation of the right-handed neutrino mass spectrum in such scenarios. 

Motivated by these facts we review in this talk alternative proposals for neutrino mass generation, through the interaction of neutrinos with non-trivial backgrounds. We examine two such types of backgrounds. The first~\cite{leite}, to be discussed in section \ref{sec:LV}, violates Lorentz symmetry spontaneously. The background is of a type existing in the so-called  Lorentz-Violating (LV) Standard Model Extension (SME) of Kostelecky and collaborators~\cite{kostel}; it can be associated (but this is only one example) with some string/brane models of the Universe, in which our four space-time dimensional brane world propagates in a bulk punctured by populations of point-like D-brane defects, interacting with right-handed neutrinos~\cite{mavromatosDfoam}. The second proposal for dynamical generation of neutrino masses concerns their propagation in space-time geometries with quantum-fluctuating torsion~\cite{mptorsion} and is discussed in section \ref{sec:torsion}.  The totally antisymmetric part of the torsion couples, via the gravitational covariant derivative, to all fermions in a way that the resulting interaction resembles that of the Lorentz and CPT-Violating pseudovector background with the axial fermion current in the SME. The generation of (right-handed, sterile) neutrino masses in that case proceeds, as we shall review below, via chiral anomalous three-loop graphs of neutrinos interacting with the totally antisymmetric torsion quantum field. In four space-time dimensions, the latter is represented as an axion field, whose mixing with ordinary axion fields, that in turn interact with the Majorana right-handed neutrinos via chirality changing Yukawa couplings, is held responsible for the right-handed Majorana neutrino mass generation through the aforementioned anomalous graphs. This latter scenario may be motivated by some 
string-inspired proposals~\cite{sarkarlepto} for leptogenesis induced by the torsion background in geometries of the early universe, which we do not discuss here, due to lack of space. The totally antisymmetric torsion in such cases corresponds to the field strength of the spin-one antisymmetric tensor (Kalb-Ramond) field of the string gravitational multiplet.  

\section{Neutrino Mass Generation due to propagation in Lorentz Violating Backgrounds \label{sec:LV}}

The model to be considered in this part of the talk is defined by the two-flavour fermion Lagrangian
\be\label{model}
\mathcal{L}_2 = \bar{\Psi} \left[i(\partial_0\gamma^0-\vec\partial\cdot\vec\gamma)\left(1-\frac{\Delta}{M^2}\right)
+\frac{\Delta}{M}\right] \Psi + \frac{1}{M^2} (\ol\Psi\tau\Psi)^2,
\ee
where
$\Psi=\begin{pmatrix} 
       \psi_1\\
       \psi_2
       \end{pmatrix}$       
is a fermion flavour doublet, with bare mass zero, 
and
\be
\tau=\begin{pmatrix} 
       g_1 \quad g_3 \\
       g_3 \quad g_2
       \end{pmatrix}~, 
\ee
is the (dimensionless) interaction coupling matrix.  The mass scale $M$ is used both to control the LV scale and the 
strength of the four-fermion interaction~\footnote{When applied to right-handed neutrinos, and in cases, such as the $\nu$MSM~\cite{nuMSM}, where  the lightest of them (of keV mass) plays the role of dark matter, the four fermion interactions provide an example of self-interacting dark matter, which may play an important r\^ole in galactic structure~\cite{ruffini}.}. 
We shall argue that in this model fermion masses and 
 flavour oscillations are generated dynamically.

The model belongs to an SME of the type 
\begin{equation}
{\cal L}_{SME}=\overline \psi\left(i\slashed\partial-m+Q\right)\psi~,
\end{equation}
where $Q$ contains the LV terms, and can be expanded on a basis of gamma matrices 
$Q=A+iB\gamma^5+C_\mu\gamma^\mu+D_\mu\gamma^\mu\gamma^5+E_{\mu\nu}\sigma^{\mu\nu},$ with $\sigma^{\mu\nu} = \frac{1}{4}[\gamma^\mu. \, \gamma^\nu]$.
The (tensorial) quantities $A,B,C_\mu,D_\mu,E_{\mu\nu}$ may contain any number of derivatives, including terms which are either odd or even under
the discrete symmetry CPT. The different coefficients associated with these terms can arise from vacuum expectation values (vev) of tensors of different ranks,
and should satisfy upper bounds for Lorentz symmetry violation, imposed by experiments \cite{kostel}. The model (\ref{model}) corresponds to the specific case
$A=\frac{b}{M}\Delta~,~~C_0=-i\frac{a}{M^2}\Delta\partial_0~,~~\vec C=-i\frac{c}{M^2}\Delta\vec\partial~,~~B=D_\mu=E_{\mu\nu}=0,$
where $a,b,c$ are dimensionless constants ($a>0$ and $c>0$), such that
$Q=-i\partial_0\gamma^0\frac{a}{M^2}\Delta+i\vec\partial\cdot\vec\gamma \left(i\frac{b}{M}\vec\partial\cdot\vec\gamma+\frac{c}{M^2}\Delta\right).$
The relevant case for our study corresponds to non-vanishing coefficients $a,b,c$, 
which leads~\cite{leite} to a quasi-relativistic dispersion relation, in the sense that it 
is relativistic in both the infrared and the ultraviolet,
but not in an intermediate regime, characterized by the mass scale $M$, and the absence of critical coupling for the generation of dynamical mass.

The choice (\ref{model}) is motivated by a gravitational microscopic model, based on the low-energy limit of a string theory on a three brane universe, which is embedded, 
from an effective three-brane observer view point, in a bulk space-time punctured with point-like defects (\emph{D-particles}) \cite{mavromatosDfoam}. 
Non-trivial interactions of D-particles with stringy matter occur only for such matter that does not carry standard model quantum numbers, and from this point of view right-handed sterile neutrinos constitute perfect candidates. 
As shown in \cite{leite}, this model provides an elegant construction of the LV operator $Q$ from a fundamental theory, 
which allows the generation of the operator $Q$ in a natural way. 
In this microscopic context, the Lorentz symmetric limit $M \to \infty$, which we shall consider in this work and by means of which one views the LV 
as a catalyst of mass generation, follows when the density of D-particles becomes vanishingly small.  

To study dynamical generation of fermion masses, we introduce a Yukawa coupling of fermions to an auxiliary field $\phi$, which is used for the linearisation of the four-fermion interactions in (\ref{model}), integrate over the 
fermions in a path integral, and look for a non-trivial minimum for the effective potential $V(\phi)$. This leads to a mass term in the original
Yukawa interaction. This approach neglects fluctuations of the auxiliary field about its vacuum expectation value (vev),  
but these can be omitted in the limit $g^2\to0$, which characterises the Lorentz symmetric limit~\footnote{In terms of the microscopic, string-inspired model~\cite{mavromatosDfoam}, the coupling is proportional to the density of D-particle defects, and hence its vanishing is consistent with 
the simultaneous limit $M \to \infty$, since $M$ is inversely proportional to the defects density. Hence the Lorentz symmetric limit of that model
corresponds to vanishing D-particle-defect density.}.

The Lagrangian containing the auxiliary field, equivalent to the original Lagrangian (\ref{model}), reads
\be\label{lag2}
\mathcal{L}_2' = \bar{\Psi}\left[i(\partial_0\gamma^0-\vec\partial\cdot\vec\gamma)\left(1-\frac{\Delta}{M^2}\right)
+\frac{\Delta}{M}\right]\Psi-\frac{M^2}{4}\phi^2-\phi\ol\Psi\tau\Psi~.
\ee
Integrating over fermions, we obtain the effective potential for the auxiliary field $\phi$~\cite{leite}  
$V_2=\frac{M^2}{4}\phi^2+i ~tr\int \frac{d^4 p}{(2 \pi)^4} (\ln\lambda_++\ln\lambda_-),$ 
where $\lambda_\pm=(\omega\gamma^0-\vec p\cdot\vec\gamma)(1+p^2/M^2)-p^2/M-h_\pm\phi,$
$\omega$ is the frequency, $p \equiv |\vec p|$, and 
$ h_\pm=\frac{1}{2}(g_1+g_2)\pm\frac{1}{2}\sqrt{(g_1-g_2)^2+4g_3^2}$ 
are the eigenvalues of the coupling matrix. 

Minimization $(dV_2/d\phi)\Big|_{\phi = \langle \phi \rangle \equiv {\phi_2}}=0$ leads to a \emph{gap} equation for the fermion masses
\be\label{gap}
\frac{M^2}{2}\phi_2=
\sum_{s=+,-}\frac{h_s}{\pi^3}\int p^2 dp \int d\omega \left[\frac{ (h_s\phi_2 + p^2/M)}{(\omega^2+p^2)(1+p^2/M^2)^2+(h_s\phi_2+p^2/M)^2} \right]~.
\ee
From (\ref{gap}) it follows that the dynamically generated mass matrix $\mathcal{M} = \phi_2 \tau$ (\emph{cf.} (\ref{lag2})) reads
in the (Lorentz-symmetric) limit of small couplings $g_i \ll 1$ we are interested in here,
\be
\mathcal{M} \simeq 0.018\, (g_1+g_2)M\begin{pmatrix} g_1 & g_3 \\ g_3 & g_2\end{pmatrix}~.
\ee
The mass eigenvalues $m_\pm = \phi_2 h_\pm$ and the mixing angle $\theta$ are then  given by
\bea\label{md12}
m_\pm &=& 0.009 \, M\, \left[(g_1+g_2)^2\pm\sqrt{(g_1^2-g_2^2)^2+4g_3^2(g_1+g_2)^2}\right]\nn
\tan\theta&=&\frac{g_1-g_2}{2g_3}+\sqrt{1+\left(\frac{g_1-g_2}{2g_3}\right)^2}~.
\eea
From this we can express the dimensionless couplings $g_i$ as
\bea\label{g1g2g3}
g_1 &=&\frac{\mu_++\mu_-+(\mu_+-\mu_-)\cos(2\theta)}{2\sqrt{\alpha(\mu_++\mu_-)}}~,\\
g_2 &=&\frac{\mu_++\mu_--(\mu_+-\mu_-)\cos(2\theta)}{2\sqrt{\alpha(\mu_++\mu_-)}}~,\nn
g_3 &=&\frac{\mu_--\mu_+}{2\sqrt{\alpha(\mu_++\mu_-)}}\sin(2\theta)~,\nonumber
\eea
where 
$ \mu_\pm = \frac{m_\pm}{M}. $
Therefore one can write the couplings $g_i$ in the form
\be\label{gi}
g_i=\frac{a_i}{\sqrt M}~~,~i=1,2,3~,
\ee
where the constants $a_i$ are kept \emph{finite} and are fixed by the ``experimental'' (in case of realistic models)  values of $m_\pm$ and $\theta$. This expression shows the explicit 
dependence of the couplings $g_i$ on the scale $M$, for the Lorentz symmetric limit $M\to\infty$ to be taken, in such a way that we are left with
two relativistic free fermions, for which flavour oscillations have been generated dynamically. Therefore any set of values for $m_\pm$ and $\theta$
can be described by the Lorentz-symmetric limit of our model, by considering the coupling constants (\ref{g1g2g3}).

The model can be straightforwardly extended to Majorana fermions, as well as three generations of fermions, including seesaw type Lagrangians,
where by a judicious choice of the appropriate couplings one can generate, in the Lorentz-symmetric limit, (right-handed) 
neutrino mass hierarchies of relevance, e.g. to $\nu$MSM and warm dark matter studies~\cite{nuMSM}.

\section{Neutrino Mass Generation due to Interactions with Quantum Torsion\label{sec:torsion}}

Let us for concreteness 
consider  Dirac fermions in  a torsionful  space-time.  The extension to the Majorana case is straightforward. 
The relevant 
action reads:
\begin{equation}\label{dirac}
S_\psi = \frac{i}{2} \int d^4 x \sqrt{-g} \Big( \overline{\psi}
\gamma^\mu \overline{\mathcal{D}}_\mu \psi  
- (\overline{\mathcal{D}}_\mu \overline{\psi} ) \gamma^\mu \psi \Big)
\end{equation}
where $\overline{\mathcal{D}}_\mu = \overline{\nabla}_\mu  + \dots $,
is the covariant derivative (including gravitational and gauge-field connection parts, in case the fermions are charged). The
overline above  the covariant derivative, i.e.~$\overline{\nabla}_\mu$,
denotes  the presence  of  torsion, which  is  introduced through  the
torsionful spin connection: $\overline{\omega}_{a  b \mu} = \omega_{a b
  \mu} + K_{a  b \mu} $, where $K_{ab \mu}$  is the contorsion tensor.
The  latter  is  related  to  the  torsion  two-form  $\textbf{T}^a  =
\textbf{d   e}^a   +    \overline{\omega}^a   \wedge   \textbf{e}^b   $
via~\cite{torsion,kaloper}:     $K_{abc}    =     \frac{1}{2}    \Big(
\textrm{T}_{cab}  - \textrm{T}_{abc}  -  \textrm{T}_{bcd} \Big)$.  The
presence  of torsion  in the  covariant derivative  in  the 
action (\ref{dirac}) leads, apart from the standard terms in manifolds
without  torsion, to an  additional term  involving the  axial current
\be\label{axial}
J^\mu_5 \equiv \overline{\psi} \gamma^\mu \gamma^5 \psi~.
\ee
The relevant part of the action reads:
\begin{equation}\label{torsionpsi}
S_\psi \ni  - \frac{3}{4} \int d^4 \sqrt{-g} \, S_\mu \overline{\psi}
\gamma^\mu \gamma^5 \psi  = - \frac{3}{4} \int S \wedge {}^\star\! J^5  
\end{equation}
where $\textbf{S} = {}^\star\! \textbf{T}$  is the dual of \textbf{T}: $S_d
=   \frac{1}{3!}     \epsilon^{abc}_{\quad   d}   T_{abc}$.     

We  next remark that  the torsion  tensor can  be decomposed  into its
irreducible parts~\cite{torsion},  of which $S_d$  is the pseudoscalar
axial vector:
$ T_{\mu\nu\rho} = \frac{1}{3} \big(T_\nu
g_{\mu\rho} - T_\rho g_{\mu\nu} \big) - \frac{1}{3!}
\epsilon_{\mu\nu\rho\sigma} \, S^\sigma + q_{\mu\nu\rho}$, 
with
$\epsilon_{\mu\nu\rho\sigma} q^{\nu\rho\sigma} = q^\nu_{\,\rho\nu} =
0$.
This implies that the contorsion tensor undergoes the following decomposition:
\begin{equation}\label{hatted}
K_{abc} = \frac{1}{2} \epsilon_{abcd} S^d + {\widehat K}_{abc} 
\end{equation}
where $\widehat  K$ includes the  trace vector $T_\mu$ and  the tensor
$q_{\mu\nu\rho}$ parts of the torsion tensor.
 
The gravitational part of the action can then be written as:
$
S_G =\frac{1}{2\kappa^2} \, \int d^4 x \sqrt{-g} \Big(R +
\widehat{\Delta} \Big) + \frac{3}{4\kappa^2} \int \textbf{S} \wedge
{}^\star\! \textbf{S},$
where  $\widehat \Delta  = {\widehat  K}^\lambda_{\ \: \mu\nu} {\widehat
  K}^{\nu\mu}_{\quad \lambda}  - {\widehat K}^{\mu\nu}_{\quad  \nu} \,
{\widehat K}^{\quad  \lambda}_{\mu\lambda}$, with the  hatted notation
defined in (\ref{hatted}).

In a  quantum gravity setting,  where one integrates over  all fields,
the torsion terms  appear as non propagating fields  and thus they can
be integrated out exactly. The authors of \cite{kaloper} have observed
though   that  the   classical  equations   of  motion   identify  the
axial-pseudovector torsion field $S_\mu$ with the axial current, since
the torsion equation yields
\begin{equation}\label{torsionec}
K_{\mu a b} = - \frac{1}{4} e^c_\mu \epsilon_{a b c d} \overline{\psi}
\gamma_5 {\tilde \gamma}^d \psi\ .
\end{equation}
From this  it follows $\textbf{d}\,{}^\star\!\textbf{S}  = 0$, leading
to a  conserved ``torsion charge'' $Q =  \int {}^\star\!  \textbf{S}$.
To  maintain  this conservation  in  quantum  theory, they  postulated
$\textbf{d}\,{}^\star\!\textbf{S} = 0$ at the quantum level, which can
be  achieved  by  the  addition  of  judicious  counter  terms.   This
constraint, in a path-integral formulation of quantum gravity, is then
implemented  via a delta  function constraint,  $\delta (d\,{}^\star\!
\mathbf{S})$, and the latter via the well-known trick of introducing a
Lagrange multiplier  field $\Phi (x)  \equiv (3/\kappa^2)^{1/2} b(x)$.
Hence, the relevant torsion  part of the quantum-gravity path integral
would include a factor {\small
\begin{eqnarray}
 \label{qtorsion}
&&\hspace{-5mm} \mathcal{Z} \propto \int D \textbf{S} \, D b   \, \exp \Big[ i \int
    \frac{3}{4\kappa^2} \textbf{S} \wedge {}^\star\! \textbf{S} -
      \frac{3}{4} \textbf{S} \wedge {}^\star\! \textbf{J}^5  +
      \Big(\frac{3}{2\kappa^2}\Big)^{1/2} \, b \, d {}^\star\! \textbf{S}
      \Big]\nonumber \\  
&&\hspace{-5mm}=\!  \int D b  \, \exp\Big[ -i \int \frac{1}{2}
      \textbf{d} b\wedge {}^\star\! \textbf{d} b + \frac{1}{f_b}\textbf{d}b 
\wedge {}^\star\! \textbf{J}^5 + \frac{1}{2f_b^2}
    \textbf{J}^5\wedge\, ^\star \textbf{J}^5 \Big]\; ,\nonumber\\
\end{eqnarray}
\hspace{-1.5mm}}
where 
$f_b = (3\kappa^2/8)^{-1/2} = \frac{M_P}{\sqrt{3\pi}}$ 
and  the  non-propagating   $\textbf{S}$  field  has  been  integrated
out. The reader  should notice that, as a  result of this integration,
the   corresponding   \emph{effective}   field   theory   contains   a
\emph{non-renormalizable} repulsive four-fermion axial current-current
interaction, characteristic of any torsionful theory~\cite{torsion}.

The torsion term, being geometrical, due to gravity, couples universally to all fermion species, not only neutrinos.
Thus, in the context of the SM of particle physics,  the axial current (\ref{axial}) is expressed as a sum over fermion species
\be\label{axial2}
J^\mu_5 \equiv \sum_{i={\rm fermion~species}} \, \overline{\psi}_i  \gamma^\mu \gamma^5 \psi_i ~.
\ee
In theories with chiral anomalies, like the quantum electrodynamics part of SM, 
the  axial current  is not
conserved at the  quantum level, due to anomalies,  but its divergence
is obtained by the one-loop result~\cite{anomalies}:
\begin{eqnarray}
   \label{anom}
\nabla_\mu J^{5\mu} \!&=&\! \frac{e^2}{8\pi^2} {F}^{\mu\nu}
  \widetilde{F}_{\mu\nu}  
- \frac{1}{192\pi^2} {R}^{\mu\nu\rho\sigma} \widetilde
{R}_{\mu\nu\rho\sigma} \nonumber\\ 
&\equiv& G(\textbf{A}, \omega)\; .
\end{eqnarray}
We  may then partially integrate  the second  term in  the exponent  on the
right-hand-side  of (\ref{qtorsion})  and take  into account (\ref{anom}).
The reader should observe that in (\ref{anom}) the torsion-free spin connection has been
used.  This can be achieved by the addition of proper counter terms in
the  action~\cite{kaloper}, which  can  convert the  anomaly from  the
initial    $G(\textbf{A},   \overline   \omega)$    to   $G(\textbf{A},
\omega)$. Using  (\ref{anom}) in (\ref{qtorsion}) one  can then obtain
for the effective torsion action in theories with chiral anomalies, such as the QED part of the SM:
\begin{equation}\label{brr}
\int D b\ \exp\Big[ - i \int \frac{1}{2}
    \textbf{d} b\wedge {}^\star\! \textbf{d} b  - \frac{1}{f_b} b
    G(\textbf{A}, \omega)  
+ \frac{1}{2f_b^2} \textbf{J}^5 \wedge \, ^\star \textbf{J}^5 \Big]\; .
\end{equation}
A concrete example of torsion  is provided by string-inspired theories, where the totally antisymmetric component $S_\mu$ of the torsion 
is identified with the field strength of the spin-one antisymmetric tensor (Kalb-Ramond (KR)~\cite{Kalb}) field
$H_{\mu\nu\rho} = \partial_{[\mu} B_{\nu\rho]}$, 
where  the  symbol  $[\dots   ]$  denotes  antisymmetrization  of  the
appropriate indices.
The string theory effective action depends only on $H_{\mu\nu\rho}$ as a consequence of the ``gauge symmetry''
$B_{\mu\nu} \rightarrow B_{\mu\nu} + \partial_{[\mu }\Theta_{\nu]} $ that characterises all string theories.
It can be shown~\cite{tseytlin} that the
terms of the effective action 
up to and including quadratic order in the Regge slope parameter $\alpha^\prime$, of relevance to the low-energy (field-theory) limit of string theory, 
which involve the H-field strength,
can  be assembled in such a way
that  only    torsionful     Christoffel    symbols, $\overline{\Gamma}^\mu_{\nu\rho}$ 
appear:
$
  \label{torsionful}
\overline{\Gamma}^\mu_{\nu\rho}\ =\ \Gamma^\mu_{\nu\rho} +
\frac{\kappa}{\sqrt{3}}\, H^\mu_{\nu\rho}\ \ne\ 
\overline{\Gamma}^\mu_{\rho\nu}\;,$
where $\Gamma^\mu_{\nu\rho}  = \Gamma^\mu_{\rho\nu}$ is  the ordinary,
torsion-free, symmetric connection,  and $\kappa$ is the gravitational
constant. In four space-time dimensions, the dual of the H-field is indeed the derivative of an axion-like field, analogous to the field $b$ above. 
For completeness we mention at this point that background geometries with (approximately) constant background $H_{ijk}$ torsion, where Latin indices denote spatial components of the four-dimensional space-time, may characterise the early universe. In such cases, the H-torsion background constitutes extra source of CP violation, necessary for lepotogenesis, and through Baryon-minus-Lepton-number (B-L) conserving processes, Baryogenesis, and thus the observed matter-antimatter asymmetry in the Universe~\cite{sarkarlepto}. Today of course any torsion background should be strongly suppressed, due to the lack of any experimental evidence for it. Scenarios as to how such cosmologies can evolve so as to guarantee the absence of any appreciable traces of torsion today can be found in \cite{sarkarlepto}.

In what follows we shall consider the effects of the quantum fluctuations of torsion, which survive the absence of any torsion background.
An important aspect of the coupling  of the torsion (or KR axion) quantum field $b(x)$ to
the  fermionic   matter  discussed   above  is  its   shift  symmetry,
characteristic of an axion field. Indeed, by shifting the field $b(x)$
by  a constant:  $b(x)  \to b(x)  +  c$, the  action (\ref{brr})  only
changes by  total derivative terms, such  as $c\, R^{\mu\nu\rho\sigma}
\widetilde{R}_{\mu\nu\rho\sigma}$               
and $c\, F^{\mu\nu}\widetilde{F}_{\mu\nu}$.  These terms are irrelevant for the
equations  of motion and  the induced  quantum dynamics,  provided the
fields fall off sufficiently fast  to zero at space-time infinity. 
The scenario for  the anomalous  Majorana mass generation  through torsion proposed in \cite{mptorsion}, and reviewed here, 
consists of augmenting the  effective action (\ref{brr}) by terms that
break such a shift symmetry. To illustrate this last point, we  first couple the KR axion $b(x)$ to
another pseudoscalar  axion field $a(x)$.   In string-inspired models,
such  pseudoscalar   axion~$a(x)$  may  be  provided   by  the  string
moduli~\cite{arvanitaki}.    The  proposed   coupling  occurs
through  a mixing  in  the kinetic  terms  of the  two  fields. To  be
specific, we consider the action (henceforth we restrict ourselves to right-handed Majorana neutrino fermion fields)
\begin{eqnarray} 
  \label{bacoupl}
    \mathcal{S} \!&=&\! \int d^4 x \sqrt{-g} \, \Big[\frac{1}{2}
      (\partial_\mu b)^2 + \frac{b(x)}{192 \pi^2 f_b}
      {R}^{\mu\nu\rho\sigma} \widetilde{R}_{\mu\nu\rho\sigma} 
      + \frac{1}{2f_b^2} J^5_\mu {J^5}^\mu + \gamma
      (\partial_\mu b )\, (\partial^\mu a ) + \frac{1}{2}
      (\partial_\mu a)^2\nonumber\\ 
&&- y_a i a\, \Big( \overline{\psi}_R^{\ C} \psi_R - \overline{\psi}_R
\psi_R^{\ C}\Big) \Big]\;, \qquad 
\end{eqnarray}
where $\psi_R^{\ C} = (\psi_R)^C$ is the charge-conjugate right-handed
fermion $\psi_R$, $J_\mu^5  = \overline{\psi} \gamma_\mu \gamma_5 \psi$
is the  axial current of  the four-component Majorana fermion  $\psi =
\psi_R  +  (\psi_R)^C$,  and  $\gamma$  is  a  real  parameter  to  be
constrained later on.   Here, we have ignored gauge  fields, which are
not of interest to us,  and the possibility of a non-perturbative mass
$M_a$  for the  pseudoscalar  field~$a(x)$.  Moreover,  we remind  the
reader that the {\em repulsive} self-interaction fermion terms are due
to the existence of torsion in the Einstein-Cartan theory.  The Yukawa
coupling $y_a$ of  the axion moduli field $a$  to right-handed sterile
neutrino  matter $\psi_R$  may  be due  to  non perturbative  effects.
These terms \emph{break} the shift symmetry: $a \to a + c$.

It is convenient to diagonalize  the axion kinetic terms by redefining
the KR axion field as follows:
$b(x) \rightarrow {b^\prime}(x) \equiv b(x) + \gamma a(x)$. 
This implies that the effective action (\ref{bacoupl}) becomes:
\begin{eqnarray} 
  \label{bacoupl2}
&& \mathcal{S} =  \int d^4 x \sqrt{-g} \, \Big[\frac{1}{2}
      (\partial_\mu b^\prime )^2 + \frac{1}{2} \Big(1- \gamma^2
\Big) \, (\partial_\mu a)^2
  \nonumber \\ && + \frac{1}{2f_b^2} J^5_\mu {J^5}^\mu  +
\frac{b^\prime (x) - \gamma a(x)}{192 \pi^2 f_b}  
{R}^{\mu\nu\rho\sigma} \widetilde{R}_{\mu\nu\rho\sigma}  - y_a i a\, \Big( \overline{\psi}_R^{\ C} \psi_R - \overline{\psi}_R
\psi_R^{\ C}\Big) \Big]\;.\qquad
\end{eqnarray}
Thus we  observe that the $b^\prime  $ field has decoupled  and can be
integrated out  in the  path integral, leaving  behind an  axion field
$a(x)$  coupled   both  to  matter   fermions  and  to   the  operator
$R^{\mu\nu\rho\sigma}   {\widetilde   R}_{\mu\nu\rho\sigma}$,  thereby
playing now the  r\^ole of the torsion field.   We observe though that
the approach is only valid for
$
|\gamma|\ <\ 1\; , $
otherwise the  axion field would  appear as a  ghost, \emph{i.e.}~with
the  wrong  sign  of  its  kinetic  terms,  which  would  indicate  an
instability  of  the  model.  This  is the  only  restriction  of  the
parameter $\gamma$. In this case we may redefine the axion field so as to appear with a
canonical normalised kinetic term, implying the effective action:
{\small
\begin{eqnarray} 
  \label{bacoupl3} 
\mathcal{S}_a \!\!&=&\!\! \int d^4 x
    \sqrt{-g} \, \Big[\frac{1}{2} (\partial_\mu a )^2 - \frac{\gamma
        a(x)}{192 \pi^2 f_b \sqrt{1 - \gamma^2}}
      {R}^{\mu\nu\rho\sigma} \widetilde{R}_{\mu\nu\rho\sigma} \nonumber\\ 
&&\hspace{-5mm} - \frac{iy_a}{\sqrt{1 - \gamma^2}} \, 
a\, \Big( \overline{\psi}_R^{\ C} \psi_R - \overline{\psi}_R
\psi_R^{\ C}\Big) + \frac{1}{2f_b^2} J^5_\mu {J^5}^\mu
      \Big]\; .
\end{eqnarray}
\hspace{-1.5mm}}
Evidently, the  action $\mathcal{S}_a$ in~(\ref{bacoupl3}) corresponds
to a  canonically normalised axion field $a(x)$,  coupled \emph{both }
to the curvature of space-time, \emph{\`a la} torsion, with a modified
coupling $\gamma/(192 \pi^2  f_b \sqrt{1-\gamma^2})$, and to fermionic
matter  with  chirality-changing  Yukawa-like  couplings of  the  form
$y_a/\sqrt{1 - \gamma^2}$.
%******************************************************************
%Figure 1:  Anomalous Majorana mass generation
%******************************************************************
\begin{figure}[t]
 \centering
  \includegraphics[clip,width=0.40\textwidth,height=0.15\textheight]{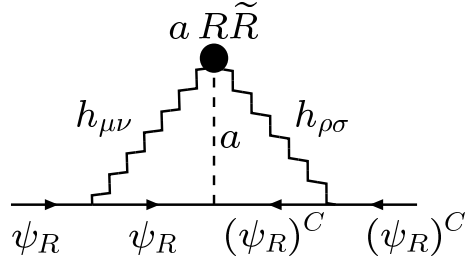} 
\caption{\it Typical Feynman graph giving rise to anomalous fermion
  mass generation~\cite{mptorsion}.  The black circle denotes the operator $a(x)\,
  R_{\mu\nu\lambda\rho}\widetilde{R}^{\mu\nu\lambda\rho}$ induced by
  torsion. The fields $h_{\mu\nu}$ denote graviton fluctuations.}\label{fig:feyn}
\end{figure}

The mechanism for  the anomalous Majorana mass generation  is shown in
Fig.~\ref{fig:feyn}.   We  may  now  estimate  the  two-loop  Majorana
neutrino mass in  quantum gravity with an effective  UV energy cut-off
$\Lambda$.    Adopting  the   effective   field-theory  framework   of
\cite{Donoghue:1994dn},  the gravitationally induced Majorana mass for right-handed neutrinos, $M_R$, is estimated to be:
\begin{equation}
  \label{MR}
M_R \sim \frac{1}{(16\pi^2)^2}\;
\frac{y_a\, \gamma\  \kappa^4 \Lambda^6}{192\pi^2 f_b (1 - \gamma^2 )} =
\frac{\sqrt{3}\, y_a\, \gamma\,  \kappa^5 \Lambda^6}{49152\sqrt{8}\,
\pi^4 (1 - \gamma^2 )}\; .  
\end{equation} 
In a UV
complete theory  such as  strings, the cutoff $\Lambda$  and the Planck mass scale $M_P$  are related.

It is interesting  to provide a numerical estimate  of the anomalously
generated Majorana mass $M_R$. Assuming that $\gamma \ll 1$, the size 
of $M_R$ may be estimated from~(\ref{MR}) to be
{\small
\begin{equation}
  \label{MRest}
M_R \sim (3.1\times 10^{11}~{\rm GeV})\bigg(\frac{y_a}{10^{-3}}\bigg)\;
\bigg(\frac{\gamma}{10^{-1}}\bigg)  
\bigg(\frac{\Lambda}{2.4 \times 10^{18}~{\rm GeV}}\bigg)^6\, .
\end{equation}
\hspace{-1.5mm}}
Obviously, the generation of $M_R$ is highly model dependent.  Taking,
for example, the quantum gravity  scale to be $\Lambda = 10^{17}$~GeV,
we  find that  $M_R$ is  at the  TeV scale,  for $y_a  =  10^{-3}$ and
$\gamma =  0.1$. However, if we  take the quantum gravity  scale to be
close  to the  GUT scale,  i.e.~$\Lambda =  10^{16}$~GeV, we  obtain a
right-handed neutrino  mass $M_R \sim  16$~keV, for the choice  $y_a =
\gamma = 10^{-3}$.   This is in the preferred  ballpark region for the
sterile   neutrino    $\psi_R$   to    qualify   as   a    warm   dark
matter~\cite{Asaka:2006ek}.

In a  string-theoretic framework, many  axions might exist  that could
mix  with each  other~\cite{arvanitaki}.  Such  a  mixing can  give rise  to reduced  UV
sensitivity of  the two-loop  graph shown in  Fig.~\ref{fig:feyn}.  In such cases, the anomalously  generated Majorana mass
may be estimated to be: \\
$M_R \sim 
\frac{\sqrt{3}\, y_a\, \gamma\,  \kappa^5 \Lambda^{6-2n} 
(\delta M^2_a)^n}{49152\sqrt{8}\, 
\pi^4 (1 - \gamma^2 )}\;$
for $n \leq 3$, and 
$ M_R \sim 
\frac{\sqrt{3}\, y_a\, \gamma\,  \kappa^5 (\delta M^{2}_a)^3}{49152\sqrt{8}\, 
\pi^4 (1 - \gamma^2 )}\; \frac{(\delta
  M^{2}_a)^{n-3}}{(M^2_a)^{n-3}}\; $
for  $n >  3$.  

It  is then  not difficult  to see  that  three axions
$a_{1,2,3}$ with next-to-neighbour mixing  as discussed above would be
sufficient  to obtain a  UV finite  result for  $M_R$ at  the two-loop
level. Of  course, beyond the two  loops, $M_R$ will  depend on higher
powers of the energy  cut-off $\Lambda$, i.e.~$\Lambda^{n> 6}$, but if
$\kappa\Lambda \ll  1$, these higher-order effects are  expected to be
subdominant.

In the above $n$-axion-mixing  scenarios, we note that the anomalously
generated  Majorana mass  term  will only  depend  on the  mass-mixing
parameters $\delta M_a^2$ of the  axion fields and not on their masses
themselves, as long as $n \le 3$. 
As a final comment we mention that the values of the Yukawa couplings $y_a$ 
may be determined by some underlying discrete symmetry~\cite{vlachos}, which for instance allows two of the 
right-handed neutrinos to be almost degenerate in mass, as required for enhanced CP violation of relevance to leptogenesis~\cite{sarkarlepto},
or in general characterises the $\nu$MSM~\cite{nuMSM}. These are interesting issues that deserve further exploration.

\end{document}